\documentclass[fleqn]{aa}       
\usepackage{graphicx}
\usepackage{amssymb}
\usepackage{amsmath}
\usepackage{subfigure}

\begin{document}

\title{Modeling nuclei of radio galaxies from VLBI radio observations}
\subtitle{Application to the BL Lac Object S5~1803+784}

\author{J. Roland\inst{1},
S. Britzen\inst{2}, N. A. Kudryavtseva\inst{2,3}\thanks{Member of the International Max Planck Research School
(IMPRS) for Radio and Infrared Astronomy at the Universities of Bonn and Cologne}, \& A. Witzel\inst{2} \& M.
Karouzos\inst{2}}

\authorrunning{Roland \& al.}
\titlerunning{Modelisation of compact radio sources}

\institute{Institut d'Astrophysique, UPMC Univ Paris 06,
           CNRS, UMR 7095, 98\,bis Bd Arago ,
           75014 Paris, France
           \and
           Max-Planck-Institut f\"ur Radioastronomie,
           Auf dem H\"ugel 69, Bonn 53121, Germany
           \and
           Astronomical Institute of St.-Petersburg State University,
           Petrodvorets, Universitetsky pr. 28, 198504 St.-Petersburg, Russia}

\offprints{J. Roland, \email{roland@iap.fr}}

\date{Received <> / Accepted <>}

\abstract{We present a new method to fit the variations of both coordinates of a VLBI component as a function of
time, assuming that the nucleus of the radio source contains a binary black hole system (BBH system). The
presence of a BBH system produces 2 perturbations of the trajectory of the ejected VLBI components. By using
only the VLBI coordinates, the problem we have to solve reduces to an astrometric problem. Knowledge of the
variations of the VLBI coordinates as a function of time contains the kinematical information, thus we are able
to deduce the inclination angle of the source and the bulk Lorentz factor of the ejected component. Generally,
there is a family of the BBH system producing the same fit to our data. To illustrate this method, we apply it
to the source 1807+784. We find that the inclination of the source is $i_{o} = 5.8^{o}\;^{+1.7}_{-1.8}$ and the
VLBI component is ejected with a bulk Lorentz factor of $\gamma = 3.7\;^{+0.3}_{-0.2}$ . We determine the family
of the BBH system which provides the best fit, assuming at first that the masses of the 2 black holes are equal
and then that the masses are different. Each family of BBH systems is characterized by $T_{p} /T_{b} \approx
1.967$, where $T_{p}$ and $T_{b}$ are the precession period of the accretion disk of the black hole ejecting the
VLBI component and the orbiting period of the BBH system.

}

\maketitle

\section{Introduction}

In previous articles, (Britzen et al. \cite{bri+01} and Lobanov \&
Roland \cite{lr05}) we have shown that VLBI and optical observations
of compact radio sources can be explained if their nuclei contain a
binary black hole system (BBH system). However, for most of the compact
radio sources we have only VLBI observations.

In this article we present a method to fit the VLBI observations
using a BBH system.

This method requires knowledge of the variations of the two
coordinates of the VLBI component as a function of time. As these
observations contain the kinematical information needed, we are able
to deduce the inclination angle of the source and the bulk Lorentz
factor of the ejected component.

We propose a geometric model that assumes that the motion of a
component ejected by the nucleus of the radio source is perturbed by two
different motions, namely:
\begin{enumerate}
    \item The precession of the accretion disk,
    \item The motion of the two black holes around the gravity center
    of the BBH system.
\end{enumerate}

In addition to these two perturbations, the ejected flow is
perturbed by the slow motion of the BBH system around the center
of gravity of the galaxy. This third perturbation is often observed
in compact radio sources and is responsible for the slow bending of
the VLBI jet, i.e., the mean ejection direction changes
slowly as the distance of the component from the core increases.
This bending becomes prominent at a distance from the core
of a few milliarcsecond ($mas$). Thus, by
studying the motion of the VLBI component
in the innermost part of the VLBI jet,
we can expect that the influence
of the slow motion of the BBH system is negligible
compared to the perturbations induced by
the precession of the accretion disk and the
motion of the black holes around the gravity center of the BBH
system. The inclusion of this slow motion
in this model would be difficult
as the characteristics of the BBH system
motion around the gravity center of the galaxy
(i.e. the radius, the speed ) are essentially unknown.

According to the above, in order to model the
ejection of VLBI components by a BBH system,
observations of a component moving in the first $mas$
of the VLBI jet are needed. Using just
the VLBI coordinates and a geometric model, the problem
we have to solve reduces to an astrometric one.

The method presented in Lobanov \& Roland (2005) was not
consistent. In this article we built a consistent method
that solves the problems found in Lobanov \& Roland (2005).

In Section 2 we will present the details of the model and in Section 3
the new method.

To illustrate the method, we apply it to the source S5~1803+784
and we show how, from the knowledge of the variations of the
coordinates $X(t)$ and $Y(t)$ of a VLBI component, we can derive
\begin{itemize}
    \item the inclination angle of the source,
    \item the bulk Lorentz factor of the ejected component,
    \item the angle between the accretion disk and the plane of the
    BBH system,
    \item the size of the BBH system,
    \item the ratio $T_{p} / T_{b}$, where $T_{p}$ is the precession
    period of the accretion disk and $T_{b}$ the orbiting period of
    the BBH system,
    \item the origin of the ejection of the VLBI component,
    \item the duration of the ejection of the plasma responsible for
    the VLBI component.
\end{itemize}

\section{The model}

\subsection{Introduction: The two-fluid model}
We will describe the ejection of a VLBI component in the framework of the
two-fluid model (Sol et al. 1989, Pelletier \& Roland 1989, 1990,
Pelletier \& Sol 1992). The two-fluid description of the outflow is
adopted with the following assumptions :
\begin{enumerate}
    \item The outflow consists of an $e^{-}-p$ plasma (hereafter
    \textit{the jet}) moving at mildly relativistic speed
    $v_{j}\leq 0.4\times c$ and an $e^{\pm}$ plasma (hereafter
    \textit{the beam}) moving at highly relativistic speed (with
    corresponding Lorentz factor $\gamma_{b}\leq 30$).
        \item The magnetic field lines are parallel to the flow in the
        beam and the mixing layer, and are toroidal in the jet
        (see Figure 1 of Lobanov \& Roland 2005).
\end{enumerate}


The $e^{-}-p$ jet carries most of the mass and the kinetic energy
ejected by the nucleus. It is responsible for the formation of
kpc-jets, hot spots and extended lobes (Muxlow et al. 1988, Roland et
al. 1988, Roland \& Hetem 1996). The relativistic $e^{\pm}$ beam
moves in a channel through the mildly relativistic jet and is
responsible for the formation of superluminal sources and their
$\gamma$-ray emission (Roland et al. 1994). The relativistic beam can
propagate if the magnetic field $B$ is parallel to the flow in the
beam and in the mixing layer between the beam and the jet and if it
is greater than a critical value (Pelletier et al. 1988, Achatz \&
Schlickeiser 1993). The magnetic field in the jet becomes rapidly
toroidal (Pelletier \& Roland 1990).

The observational evidence for the
two-fluid model has been discussed by e.g. Roland \& Hetem (1996).
Recent observational evidence for relativistic ejection of
an $e^{\pm}$ beam come from the $\gamma$-ray observations of MeV
sources (Roland \& Hermsen 1995, Skibo et al. 1997) and from VLBI
polarization observations (Attridge et al. 1999).

The formation of X-ray and $\gamma$-ray spectra, assuming relativistic
ejection of $e^{\pm}$ beams, has been investigated by Marcowith et
al. (1995, 1998) in the case of Centaurus A.

The possible existence of
VLBI components with two different speeds has been recently pointed
out in the case of the radio galaxies Centaurus A (Tingay et
al. 1998), Virgo A (Biretta et al. 1999) and 3C 120 (Gomez et al. 2001).
If the relativistic beam transfers some energy and/or relativistic particles
to the jet, the relativistic particles in the jet will radiate and
a new VLBI component with a mildly relativistic speed will be observed
(3C 120 is a good example of a source showing this effect).


\subsection{The geometry of the model}

\begin{figure}[ht]
\centerline{
\includegraphics[scale=0.3, bb =-100 -50 700 600,clip=true]{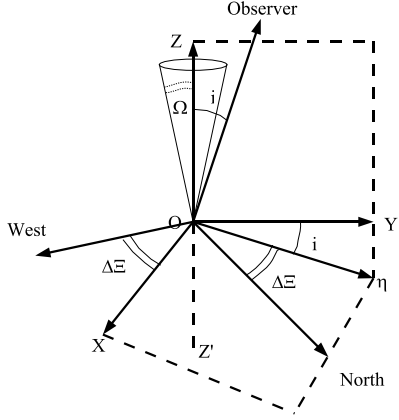}}
\caption{The geometry of the problem}
\label{fig:Frame}
\end{figure}

We will call $\Omega$ the angle between the accretion disk and the orbital plane
($XOY$) of the BBH system. The ejection of the component
will be on a cone with its axis in the $Z'OZ$ direction and opening angle
$\Omega$. We will assume that the line of sight is in
the plane ($YOZ$) and forms an angle $i_{o}$ with the axis $Z'OZ$
(see Figure \ref{fig:Frame}). The plane perpendicular to the line of
sight is the plane ($\eta OX$). We will call $\Delta\Xi$ the angle
of the rotation in the plane perpendicular to the line of sight, in order to transform
the coordinates $\eta$ and $X$ into coordinates $N$ (North) and $W$ (West), which are directly comparable with the VLBI observations. We have

\begin{equation}
W  =  x \cos(\Delta\Xi) - (z \sin(i_{o}) + y \cos(i_{o})) \sin(\Delta\Xi)\ ,
\label{eq:West}
\end{equation}

\begin{equation}
N  =  x \sin(\Delta\Xi) + (z \sin(i_{o}) + y \cos(i_{o})) \cos(\Delta\Xi)\ .
\label{eq:North}
\end{equation}

\subsection{The precession model}

Here we describe the precession of the accretion disk.

The coordinates of a component moving in the perturbed beam are given by

\begin{equation}
x_{c}  = R_{o}(z)cos(\omega_{p}t-k_{p}z(t)+\phi_{o})\ ,
\label{eq:x}
\end{equation}

\begin{equation}
y_{c} = R_{o}(z)sin(\omega_{p}t-k_{p}z(t)+\phi_{o})\ ,
\label{eq:y}
\end{equation}

\begin{equation}
z_{c} =  z_{c}(t)\ ,
\label{eq:z}
\end{equation}
where $\omega_{p} = 2 \pi / T_{p}$, $T_{p}$ is the precession period, and
$k_{p}$ is defined by

\begin{equation}
    k_{p} = 2 \pi/T_{p}V_{a}\ ,
\end{equation}
where $V_{a}$ is the speed of the propagation of the perturbations and a
free parameter of the problem.

We will assume that the amplitude of the perturbation first increases linearly,
and we take the form of the amplitude $R(z_c(t))$ to be

\begin{equation}
    R(z_{c}(t)) = \frac{R_{o}z_{c}(t)}{(a +z_{c}(t))}\ ,
\label{eq:Ro}
\end{equation}
where $a$ is
\begin{equation}
a = R_{o}/(2\;tan\Omega)\ .
\label{eq:Omega}
\end{equation}

\subsection{The binary system model}
To explain the origin of the precession of the accretion disk,
we propose that the nucleus hosts a BBH system.

As stated above, the two black holes orbit in the plane (XOY),
and the origin of our coordinate system is centered on the mass center of the system.
The elliptical orbit is given by

\begin{equation}
    r = \frac{p}{1 + e\;cos(\varphi)}\ ,
\end{equation}
where $e$ and $p$ are respectively the eccentricity and the
parameter or the semi-latus rectum of the orbit.

We will assume that the two black holes have circular orbits,
i.e. $e=0$ and we will define the black hole that ejects the VLBI
component with index 1. Its coordinates are :

\begin{equation}
    X_{1}(t) = -\frac{M_{2}}{M_{1}+M_{2}} p \cos(\psi(t))\ ,
\end{equation}

\begin{equation}
    Y_{1}(t) = -\frac{M_{2}}{M_{1}+M_{2}} p \sin(\psi(t))\ .
\end{equation}

As the orbits are circular, we have $\psi(t) = \omega_{b}t +
\psi_{o}$. Writing
\begin{equation}
    x_{1} = y_{1} = -\frac{p M_{2}}{M_{1}+M_{2}} \ ,
\end{equation}
we have
\begin{eqnarray}
    x_{1} = -\frac{ M_{2}}{M_{1}+M_{2}} \times \left[\frac{T_{b}^{2}}{4\pi^{2}}G(M_{1}+M_{2}) \right]^{1/3} \ ,
\end{eqnarray}
where $T_{b}$ is the period of the BBH system.

We define $R_{bbh}$ the distance between the two black holes as
the size of the BBH system, it is:

\begin{eqnarray}
    R_{bbh} = \left[\frac{T_{b}^{2}}{4\pi^{2}}G(M_{1}+M_{2}) \right]^{1/3} \ .
\end{eqnarray}

The coordinates of black hole 1 can be written

\begin{equation}
    X_{1}(t) = x_{1}\cos(\omega_{b}t + \psi_{o})\ ,
\end{equation}

\begin{equation}
    Y_{1}(t) = y_{1}\sin(\omega_{b}t + \psi_{o})\ .
\end{equation}






Finally, we suppose that the perturbation of the beam
is damped with a characteristic time $T_{d}$.

For VLBI observations the origin of the coordinates is
black hole 1, i.e. the black hole ejecting the VLBI components.
Therefore, the coordinates of the moving components \textit{in the frame
of reference where black hole 1 is considered the origin} are

\begin{eqnarray}
x_{c} & = & [R_{o}(z) \cos(\omega_{p}t-k_{p}z(t)+\phi_{o}) \nonumber \\
        &&{} + x_{1} cos(\omega_{b}t-k_{b}z(t)+\psi_{o}) - x_{1} \cos(\psi_{o})] \nonumber \\
        &&{} exp(-t/T_{d})\ ,
\label{eq:xc}
\end{eqnarray}

\begin{eqnarray}
y_{c} & = & [R_{o}(z) \sin(\omega_{p}t-k_{p}z(t)+\phi_{o}) \nonumber \\
        &&{} + y_{1} \sin(\omega_{b}t-k_{b}z(t)+\psi_{o}) - y_{1} \sin(\psi_{o})] \nonumber \\
        &&{} exp(-t/T_{d})\ ,
\label{eq:yc}
\end{eqnarray}

\begin{equation}
z_{c} = z_{c}(t)\ ,
\label{eq:zc}
\end{equation}
where $\omega_{b} = 2 \pi/T_{b}$,
and $k_{b}$ is defined by

\begin{equation}
    k_{b} = \frac{2 \pi}{T_{b} V_{a}} .
\end{equation}

The differential equation governing the evolution of
$z_{c}(t)$ can be obtained through the relation for the speed
of the component

\begin{equation}
    v_{c}^{2} = \left(\frac{dx_{c}(t)}{dt}\right)^{2} + \left(\frac{dy_{c}(t)}{dt}\right)^{2} +
    \left(\frac{dz_{c}(t)}{dt}\right)^{2}\ ,
\label{eq:v2}
\end{equation}
where $v_{c}$ is related to the bulk Lorentz factor by
$v_{c}/c = \sqrt{(1 - 1/\gamma_{c}^{2})}$.

Using (\ref{eq:x}), (\ref{eq:y}) and (\ref{eq:z}), we find from
(\ref{eq:v2}) that $dz_c/dt$ is the solution of the equation

\begin{equation}
    A\left(\frac{dz_{c}}{dt}\right)^{2} + B\left(\frac{dz_{c}}{dt}\right) + C = 0\ .
\label{eq:dzt}
\end{equation}

A calculation of the coefficients $A$, $B$ and $C$ can be found
in Appendix I.

Equation (\ref{eq:dzt}) admits two solutions corresponding to
the jet and the counter-jet.

We assumed that the line of sight is in the plane (YOZ) and
makes an angle $i_{o}$ with the $z$ axis (see Figure \ref{fig:Frame}).
Thus following Camenzind \& Krockenberger (1992), if we call
$\theta$ the angle between the velocity of the component and
the line of sight we have

\begin{equation}
    cos(\theta(t))=\left(\frac{dy_{c}}{dt}sin\;i_{o}+\frac{dz_{c}}{dt}cos\;i_{o}\right)/v_{c}\ .
    \label{eq:cos}
\end{equation}

The Doppler beaming factor $\delta$, characterizing the
anisotropic emission of the moving component, is
\begin{equation}
    \delta_{c}(t) = \frac{1}{\gamma_{c} \left[1 - \beta_{c} cos(\theta(t))\right]}\ ,
    \label{eq:delta}
\end{equation}
where $\beta_{c} = v_{c} /c$. The observed flux density is

\begin{equation}
    S_{c} = \frac{1}{D^{2}}\delta_{c}(t)^{2+\alpha_{r}}(1+z)^{1-\alpha_{r}}\int_{c}j_{c}dV\ ,
    \label{eq:flux}
\end{equation}
where $D$ is the luminosity distance of the source, $z$ its redshift,
$j_{c}$ is the emissivity of the component, and $\alpha_{r}$
is the synchrotron spectral index (a negative definition of
the spectral index, $S\propto\nu^{-\alpha}$ is used). As
the component is moving relativistically toward the observer,
the observed time is shortened and is given by

\begin{equation}
    t_{obs} = \int_{0}^{t}\left[1- \beta_{c} cos(\theta(t'))\right]\left(1+z\right)dt'\ .
    \label{eq:tobs}
\end{equation}

\section{The global method}
\label{sc:sect3}
\subsection{Introduction}
In Lobanov \& Roland (2005) we provide a method to determine the
characteristic parameters of the BBH system using radio and optical
observations. The method consists of two steps, i.e. in a
first step we use VLBI observations to model the precession (without a
BBH system) and in a second step we use optical observations to
obtain the characteristic parameters of the BBH system.

The above described method has the following problems:
\begin{enumerate}
    \item Using a simple precession model, we find for 3C 345
    that a bulk Lorentz factor increasing with time is necessary.
    \item \textit{Using the parameters of the precession found in the
    first step,} the BBH solution obtained in the second step
    is not necessarily consistent with the precession solution
    found previously. Indeed, in the limit
    $M_{2}\rightarrow 0$ the BBH solution is not
    necessarily able to reproduce the results of the precession model
    found in the first step. The limit $M_{2}\rightarrow 0$ corresponds
    to a single black hole and the precession of the accretion disk is
    due to the Lens-Thirring effect in that case.
\end{enumerate}
The above problems can be solved if we directly model the VLBI observations
with a BBH system instead of a simple precession model. In the BBH
system model, the bulk Lorentz factor is constant, and the model
explains the apparent variations of the speed of the VLBI
component when it escapes from the nucleus (see Figure \ref{fig:V_ap-D_c}; the
apparent speed of the ejected component changes by a factor of four with a constant
bulk Lorentz factor and it is not necessary to involve any acceleration
or decceleration of the VLBI component). Since the BBH system and the
precession solution are obtained simultaneously, they are obviously
self-consistent.

We call the method we present in this article the global method.

In the global method, we calculate the projected trajectory on the plane
of the sky of a component ejected by a BBH system and we determine the
parameters of the model to simultaneously produce the best fit for both the
West and North coordinates, i.e. $W_{c}(t)$, $N_{c}(t)$.
So, the parameters found are such that minimize
\begin{equation}
    \chi^{2}_{t}= \chi^{2}(W_{c}(t)) + \chi^{2}(N_{c}(t)) \ ,
    \label{eq:chi2}
\end{equation}
where $\chi^{2}(W_{c}(t))$ and $\chi^{2}(N_{c}(t))$ are the
$\chi^{2}$ calculated by comparing the VLBI observations with the
calculated coordinates $W_{c}(t)$ and $N_{c}(t)$ of the component.

\subsection{The coordinates of the VLBI component}
Solving (\ref{eq:dzt}), we determine the coordinate $z_{c}(t)$ of a
point source component ejected relativistically in the perturbed beam.
Then, using (\ref{eq:xc}) and (\ref{eq:yc}), we can find the coordinates
$x_{c}(t)$ and $y_{c}(t)$ of the component. In addition, for each point
of the trajectory, we can calculate the derivatives  $dx_{c}/dt$, $dy_{c}/dt$,
$dz_{c}/dt$ and then deduce $\cos\theta$ from (\ref{eq:cos}), $\delta_{c}$
from (\ref{eq:delta}), $S_{\nu}$ from (\ref{eq:flux}) and $t_{obs}$
from (\ref{eq:tobs}).

When the coordinates $x_{c}(t)$, $y_{c}(t)$ and $z_{c}(t)$ have been
calculated, they can be transformed to $w_{c}(t)$ (West) and
$n_{c}(t)$ (North) coordinates using (\ref{eq:West}) and
(\ref{eq:North}).

As explained in Britzen et al. (\cite{bri+01}) and Lobanov \& Roland (2005),
the radio VLBI component has to be described as an extended component along the beam.
Let us call $n_{rad}$ the number of points (or steps along the beam)
for which we integrate, in order to model the component. The coordinates
$W_{c}(t)$, $N_{c}(t)$ of the VLBI component are then
\begin{equation}
    W_{c}(t) = \left(\sum_{i=1}^{n_{rad}} w_{ci}(t)\right)/n_{rad} \ ,
\end{equation}
\begin{equation}
    N_{c}(t) = \left(\sum_{i=1}^{n_{rad}} n_{ci}(t)\right)/n_{rad} \ .
\end{equation}
and can be compared with the observed coordinates of the VLBI component.

\subsection{The parameters of the model}

In this section, we list what \textit{a priori} the free
parameters of the model are:

\begin{itemize}
  \item $i_{o}$ the inclination angle,
    \item $\phi_{o}$ the phase of the precession at $t=0$,
    \item $\Delta\Xi$ the rotation angle in the plane perpendicular
    to the line of sight (see (\ref{eq:West}) and (\ref{eq:North})),
    \item $\Omega$ the opening angle of the precession cone (see (\ref{eq:Omega})),
    \item $R_{o}$ the maximum amplitude of the perturbation (see (\ref{eq:Ro})),
    \item $T_{p}$ the precession period of the accretion disk,
    \item $T_{d}$ the characteristic time for the damping of the beam perturbation,
    \item $M_{1}$ the mass of the black hole ejecting the radio jet,
    \item $M_{2}$ the mass of the secondary black hole,
    \item $\gamma_{c}$ the bulk Lorentz factor of the VLBI component,
    \item $\psi_{o}$ the phase of the BBH system at $t=0$,
    \item $T_{b}$ the period of the BBH system,
    \item $t_{o}$ the origin of the ejection of the VLBI component,
    \item $V_{a}$ the propagation speed of the perturbations,
    \item $n_{rad}$ is the number of steps to describe the extension of the VLBI component
    along the beam.
\end{itemize}

To begin with, we assume that $M_{1} = M_{2}$ and when the corresponding solution
is obtained, we calculate the family $M_{1}(M_{2})$ which provides the same fit.

So, the problem we have to solve is a 14 free parameters problem.

If, in addition to the radio, optical observations are available
that peak in the light curve, this optical emission can be modelled
as the synchrotron emission of a point source
ejected in the perturbed beam (Britzen \& al. \cite{bri+01} and
Lobanov \& Roland 2005). This short burst of very energetic
relativistic $e^{\pm}$ is followed immediately by a very long burst
of less energetic relativistic $e^{\pm}$. This long burst is
modelled as an extended structure along the beam and is responsible for
the VLBI radio emission. In that case the origin $t_{o}$ of the VLBI
component is the beginning of the first peak of the optical
light curve and is not a free parameter of the model.

We have to investigate the different possible scenarios with regard to the
sense of the rotation of the accretion disk and the sense of the orbital
rotation of the BBH system. These possibilities correspond to $\pm\:
\omega_{p}(t- z/V_{a})$ and $\pm \:\omega_{b}(t- z/V_{a})$. As the
sense of the precession is always opposite to the sense of the
orbital motion, we will study the two cases denoted by $+-$ and $-+$
where we have $\omega_{p}(t- z/V_{a})$, $-\omega_{b}(t- z/V_{a})$ and
$-\omega_{p}(t- z/V_{a})$, $\omega_{b}(t- z/V_{a})$ respectively.

\subsection{The method to solve the problem}
To find a solution for the above described problem
we use the following method.

As mentioned before we start with the assumption that
$M_{1} = M_{2}$, i.e. that the two masses of the BBH system are equal.

First, we find the inclination angle that provides the best fit.
To do that we minimize $\chi^{2}_{t}(i_{o})$ (see (\ref{eq:chi2}))
when the inclination angle varies gradually between
two values. At each step of $i_{o}$, we determine each
free parameter $\lambda$ such that $\partial \chi^{2}_{t}/\partial \lambda = 0$
and $\chi^{2}_{t}$ presents a minimum.

Furthermore, using the inclination angle determined previously,
we explore the space of the solutions, for a varying mass
of the BBH system (while still assuming $M_{1} = M_{2}$). This
allow us to find whether the solution of the BBH system
presents a degeneration or if there are other solutions,
with different masses, that fit the observations. When
exploring the solutions space, we always vary
one parameter in a step-wise manner, with each step minimizing
$\chi^{2}_{t}$ for each of the free parameters.

The space of the solutions can be explored for each of
the free parameters if necessary.

Because the problem is a non-linear one, we calculate again
the variations of $\chi^{2}_{t}(i_{o})$ for the best solution
found previously, starting from the inclination angle obtained
in the first step. Where $\chi^{2}_{t}(i_{o})$ reaches its
minimum, we have

\begin{eqnarray}
    \left(\frac{\partial \chi^{2}}{\partial i_{o}}\right)_{min} = A\;(i_{o} - i_{o,min}) \ ,
\end{eqnarray}
and the $1\;\sigma$ error bar, $(\Delta i_{o})_{1 \sigma}$, corresponding to the
parameter $i_{o}$ is then given by

\begin{eqnarray}
    (\Delta i_{o})_{1 \sigma} = 1/A \ .
\end{eqnarray}

This assumes that around the minimum, $\chi^{2}_{t}(i_{o})$
is a parabola, however, for large variations of $i_{o}$, the
parabola approximation is not valid and a better determination
of the $1\;\sigma$ error bar can be obtained using the definition

\begin{eqnarray}
    (\Delta i_{o})_{1 \sigma} = | i_{o}(\chi^{2}_{min} + 1) -  i_{o}(\chi^{2}_{min}) | \ ,
    \label{eq:1_sigma}
\end{eqnarray}
which provides two values $(\Delta i_{o})_{1 \sigma+}$ and
$(\Delta i_{o})_{1 \sigma-}$ (see Lampton et al. 1976 and
H\'ebrard \& al. 2002).

Because we calculated $\chi^{2}_{t}(i_{o})$ by
minimizing $\chi^{2}_{t}(\lambda)$ at each step and
for each free parameter $\lambda$, we can deduce the range of
values corresponding to $1\;\sigma$ for each parameter when
the inclination varies between
$i_{o} = i_{o,min} - (\Delta i_{o})_{1 \sigma-}$ and
$i_{o} = i_{o,min} + (\Delta i_{o})_{1 \sigma+}$ .

Finally, when the best solution corresponding to $M_{1} = M_{2}$ is
obtained, we can determine the family of BBH systems with $M_{1} \neq M_{2}$
that provides the same fit.

One important point is that
using this phenomenological method, we do not have the proof
that the minimum found is unique, since for a completely
different set of values for the parameters of the problem,
another minimum could exist. However, as we will explore
a wide range of inclination angles, i.e. $1^{o} \leq i_{o}\leq 25^{o}$
and a equally wide range of BBH system masses,
i.e. $10^{6}\times M_{\odot} \leq M \leq 10^{11}\times M_{\odot}$,
we minimize the possibility of missing the best solution.
Another way to overcome this difficulty is to explore the
space of possible values of the parameters using, for instance,
a Monte Carlo Markov chain algorithm (MCMC algorithm). However this is
out of the scope of this article.

\section{Application to S5~1803+784}

\subsection{The radio morphology of S5~1803+784}

The blazar S5~1803+784 ($z\approx0.68$, Lawrence et al. 1987,
Stickel et al. 1993) is an intraday-variable (IDV) source with rapid
flux-density variations in the optical and radio regime (Wagner \&
Witzel 1995) on timescales as short as 50 minutes in the optical
(Wagner et al. 1990). The source has been observed and studied with different
VLBI arrays at a wide range of frequencies (e.g., Eckart et al.
1986, 1987; Witzel et al. 1988; Charlot 1990; Strom \& Biermann 1991;
Fey et al. 1996; Gabuzda 1999; Gabuzda \& Cawthorne 2000; Gabuzda \&
Chernetskii 2003; Ros et al. 2000, 2001; Britzen et al. 2005a,
2005b). Global VLBI observations reveal the pc-scale jet of
S5~1803+784 to be oriented in the East-West direction (e.g., Britzen
et al. 2005b), while the large-scale structure comprises a dominant
core component and a weak secondary component $\sim 45''$ away from
the core at position angle $\Theta\approx-166^{\circ}$
south-south-west (Antonucci et al. 1986; Britzen et al. 2005a) and
makes this source a "misaligned" object (e.g., Pearson \& Readhead
1988). Strom \& Biermann (1991) also detected this secondary
component, as well as a bridge connecting it with the core,
at 1.5 GHz observations with the Westerbork Synthesis Radio
Telescope (WSRT). World-array VLBI observations revealed that
the jet bends by around $90^\circ$ at a core separation of about 0.5
arcseconds towards the South (Britzen et al. 2005a). The Westerbork
Synthesis Radio Telescope (WSRT) observations resolved the bridge
connecting the core with the secondary component some $45''$ to the
south. Wiggles in the ridge of this emission suggest that, as on the
100-pc scale, the jet may also oscillate on 50-100 kpc scales. In
addition, amorphous emission to the north, as well as an extended,
halo-like component around the nucleus has been detected (Britzen et
al. 2005a).

\begin{figure}[ht]
\centerline{
\includegraphics[scale=0.36, bb =-10 -15 750 550,clip=true]{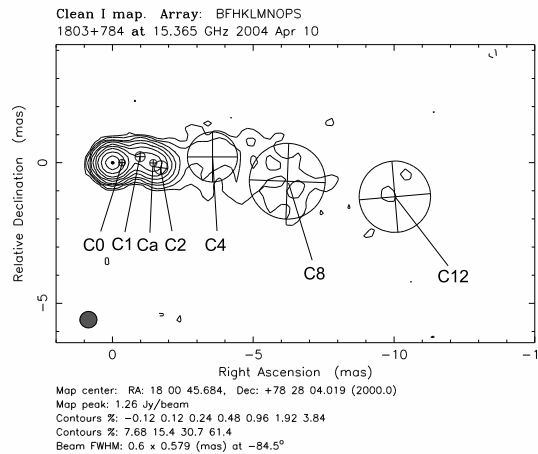}}
\caption{The pc-scale structure of 1803+784 as obtained in 15 GHz VLBA observations
in April 2004. The jet components up to a core separation of $\sim$12 mas are marked.}
\label{identification}
\end{figure}

On pc-scales the source shows a pronounced jet, with prominent jet
components located at distances between 1.4 and 12 $mas$ from the
core (e.g., Eckart et al. 1986). Geodetic and astronomical VLBI data
obtained at various epochs between 1979 and 1985 indicated that the
component at 1.4 $mas$ is stationary (e.g., Schalinski et al. 1988;
Witzel et al. 1988). Britzen et al. (2005b) discuss significant
position shifts for the 1.4 $mas$ component with displacements
between $r \sim 0.7$ $mas$ and $r \sim 1.5$ $mas$. Obviously the
improved time sampling of the geodetic VLBI data led to the detection
of systematic position variations for this component regarded as
stationary on the basis of the less frequently targeted astronomical
observations. This oscillatory behaviour is explained as the result
of a reconfinement shock (Britzen et al. 2005b). A recent
analysis of 94 epochs of VLBI observations obtained at
$\nu$ = 1.6, 2.3, 5, 8 15, and 43 GHz reveals that that the jet
structure within 12 mas from the core can most easily be described by
four jet components that remain at similar core separations in
addition to the already known ``oscillating'' jet feature at $\sim$
1.4 mas. We show the pc-scale structure of 1803+784 in an image
obtained in VLBA observations performed at 15 GHz (April 2004)
in Fig.~\ref{identification}. The jet components are indicated.
In addition to these ``stationary'' components, we find one much
fainter component, B3, moving with apparent superluminal velocities.
This component most likely coincides with the component seen in 43
GHz maps by Jorstad et al. (2005). However, at 15 GHz this component
is much fainter compared to the other components.

Evidence is presented for
quasi-periodicities in the variability of the core separation, the
position angle and the flux of the "oscillating" jet components on
timescales comparable to those derived from the total flux density
changes (Britzen et al., in prep.; Kudryavtseva et al., in prep.).
Superluminal motion has been detected in the jet of S5~1803+784 at
43, 22, (Krichbaum et al. 1993) and 8.4 GHz (Britzen et al. 2005b;
Britzen et al., in prep.). In Britzen et al. (2005b) we find and
discuss that three jet components approach the brightest and so-called
``stationary'' component (at $\sim$ 1.4 mas at 8.4 GHz) with an apparent
superluminal motion of 8\:--\:11 $c$. In this paper we show that the ``stationary''
component oscillates, -- under the assumption that this is the brightest jet
component in each epoch --  and we discuss several possible
explanations. Based on the most recently performed investigations of
a much larger database (94 epochs of VLBI observations) covering a
longer time span (of almost 20 years) and several frequencies
(1.6, 2.3, 5, 8.4, and 15 GHz) Britzen et al. find that the brightest
jet components near the core do not reveal fast radial motion.
Instead, components tend to remain at similar core separations but
show significant motion with regard to the position angle.
Some indication of apparent superluminal
motion between $\sim 5\; c$ and $7 \;c$ for the 25 $mas$ jet
component has been derived from 6 and 18 cm observations (Britzen et
al. 2005a). VLBI observations at 43 GHz revealed, for the first time,
evidence for a jet structure described as "helical" by Krichbaum (e.g.,
1990). A curved-jet morphology is found at all scales investigated so
far (e.g., Britzen et al. 2005a).

\subsection{ The observations}

For the present analysis we use the archival VLBI observations at two
frequencies, 8 and 15 GHz, spanning almost 20 years of
observations (for details see Britzen et al., in prep.). S5~1803+784
has been observed with VLBI at $\lambda$ = 15 GHz by P\'erez-Torres
et al. 2000, Kellermann et al. (1998, 2004), and Zensus et al. (2002)
between 1994.67 and 2005.68 in 13 epochs. The 8 GHz observations were
performed by Ros et al. (2000, 2001) and P\'erez-Torres et al. (2000)
from 1986.21 to 1993.95 in 41 epochs. The data had been fringe-fitted
and calibrated before by the individual observers (for details,
see the references given in Britzen et al., in prep.). We
performed modelfitting of circular Gaussian components within the
{\it Difmap} package. In order to find the optimum set of components
and parameters, we fitted each data set starting from a point-like
model. Circular components have been chosen in order to simplify the
comparison and to avoid unlikely and extremely extended elliptical
components. For details of the data reduction and the modelfit
parameters see to Britzen et al. (in prep.).

\subsection{ Preliminary remarks on the fit}
Before we begin the fit of VLBI components of S5~1803+784, we
indicate some values we will use for the cosmological
parameters and the error bars.

The cosmological model we will use is defined by
$\Omega_{t} = \Omega_{\Lambda}+ \Omega_{m}$, with
$\Omega_{\Lambda} = 0.7$, $\Omega_{m} = 0.3$ and $H_{o} = 71$ km/s/Mpc
for the Hubble constant. $\Omega_{t}$, $\Omega_{\Lambda}$ and
$\Omega_{m}$ are respectively the total density of the Universe,
the density of the vacuum and the density of the matter.
Calling $z_{s}$ the redshift of the source, the luminosity
distance, $D_{l}$, is defined by

\begin{eqnarray}
    D_{l} = \frac{1 + z_{s}}{H_{o}}\; c \int_{0}^{z_{s}}\frac{dz'}{E(z')} \ ,
\end{eqnarray}
where

\begin{eqnarray}
    E(z_{s}) = \sqrt{0.3(1 + z_{s})^{3}+0.7(1+z_{s})} \ .
\end{eqnarray}

With $z_{s} \approx 0.68$ we have $D_{l} \approx  c \times 0.8995/Ho$
and the angular distance is $D_{a} = D_{l} /(1+z)^2$.

The beams corresponding to 8 GHz and 15 GHz observations
are respectively $\approx 1.0$ $mas$ and $\approx 0.5$ $mas$.
The minimum error bars we will use are $\Delta w = 0.08$ $mas$
and $\Delta n = 0.08$ $mas$ for the West and North coordinates
at 8 GHz and $\Delta w = 0.04$ $mas$ and $\Delta n = 0.04$ $mas$
for the West and North coordinates at 15 GHz.


\subsection{VLBI components C0, C1 and the trajectory of the ejected plasma
component}

\begin{figure}[ht]
\centerline{
\includegraphics[scale=0.43, clip=true]{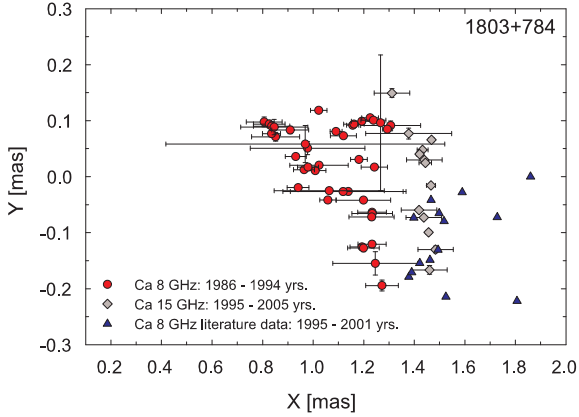}}
\caption{The movement of the Ca component. The red circles represent
the position in rectangular coordinates of the Ca component at 8 GHz
during the period 1986 -- 1994. The blue triangles show the position
at 8 GHz during 1995 -- 2001 and the grey diamonds the
position at 15 GHz from 1995 to 2005. The component moved 0.4 mas
during ten years of observations.} \label{fig:Camov}
\end{figure}

\begin{figure}[ht]
\centerline{
\includegraphics[scale=0.5, bb =-200 -20 700 350,clip=true]{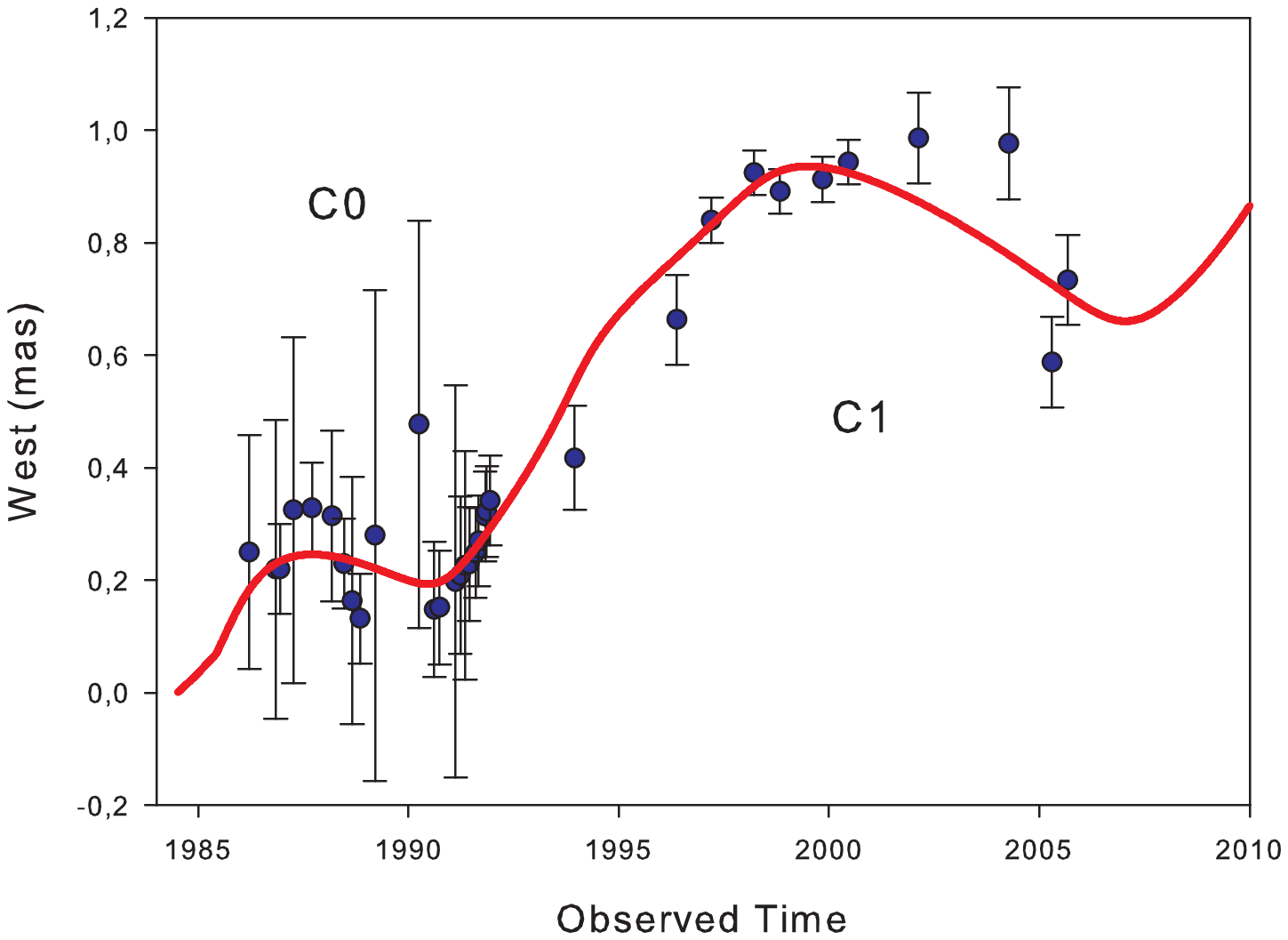}}
\caption{The West coordinate of $S5~1803+784$. The smallest error
bars for 8 GHz observations are 0.08 $mas$ and for 15 GHz observations
are 0.04 $mas$} \label{fig:Xt}
\end{figure}

\begin{figure}[ht]
\centerline{
\includegraphics[scale=0.5, bb =-200 -20 700 350,clip=true]{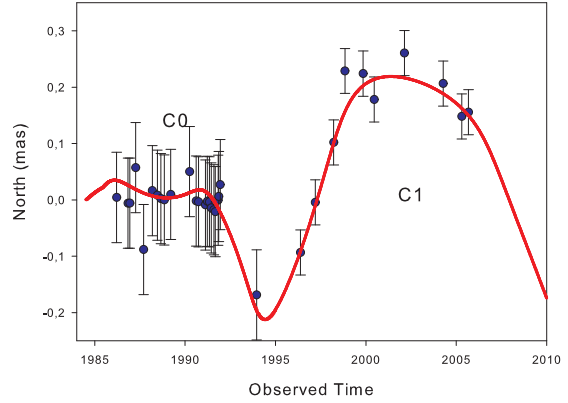}}
\caption{The North coordinate of $S5~1803+784$. The smallest error
bars for 8 GHz observations are 0.08 $mas$ and for 15 GHz observations
are 0.04 $mas$} \label{fig:Yt}
\end{figure}

\begin{figure}[ht]
\centerline{
\includegraphics[scale=0.5, bb =-200 -20 700 350,clip=true]{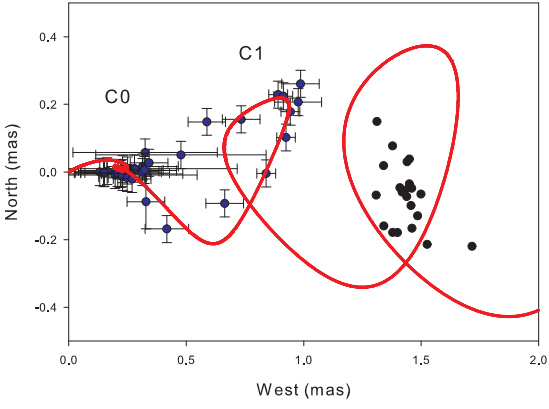}}
\caption{The trajectory of $S5~1803+784$. In addition to components
C0 and C1, we plotted the points corresponding to the 20 years of
observations of the component Ca. The solid line shows the
fit of a model to the trajectories of C0 and C1 components.
The position of the Ca component lies on the
fitted trajectory. This shows the slow motion of the beam
perturbation which is due to the perturbation speed $V_{a} \ll c$ and
the slow motion of the BBH system around the gravity center of the
galaxy.} \label{fig:Traj}
\end{figure}

The jet of S5~1803+784 consists of five "oscillating" jet
components, C0 at $\sim 0.3$ mas, C1 at $\sim 0.7$ mas, Ca at
$\sim 1.4$ mas, C2 at $\sim 2$ mas and C4 at about $\sim 4$ mas (for
the details see Britzen et al. in prep.). In this paper we will
discuss only the motion of the two innermost components C0 and C1.
Previous analysis has shown that the brightest component Ca, which is
quasi-stationary and has almost the same core separation,
is changing its position very slowly with a
speed of about 0.04 mas/yr (see Fig.~\ref{fig:Camov}).
This could mean that all the oscillating jet
components of S5~1803+784 are moving extremely slowly outwards with
time. The pertubation propagation speed of the magnetic tube (the
red curve of Figure \ref{fig:Traj}) is $V_{a} \ll c$ and is the same for
the magnetic loops corresponding to C0, C1 and CA.
The source S5~1803+784 has been observed over a time span of
20 years. Thus, we can trace this slow motion of the components with time.
The distance from the core of the component C0 is $\sim 0.3$ mas and
given a speed of about 0.04 mas/yr it will change its position by
$\sim 0.7$ mas in 10 years, which is actually the position of the
component C1. Therefore, we can assume that the component C0, which
is observed as C0 during the first six years of observations, (between 1986
and 1992), becomes component C1 and from epoch 1996.38
follows the trajectory of this component. So, by using the 20 years of
VLBI observations of S5~1803+784 as follows:
\begin{enumerate}
    \item from 1986.21 to 1991.94, VLBI data corresponding to C0 (8 GHz),
    \item for 1993.95, the VLBI point located between C0 and C1 (8 GHz), and
    \item from 1996.38 to 2005.68, VLBI data corresponding to C1 (15 GHz),
\end{enumerate}
we can calculate the trajectory that this new plasma component follows
and define the moment of ejection from the core, which we find to be around epoch
1984.5 (see the Figures \ref{fig:Xt} , \ref{fig:Yt} and \ref{fig:Traj}).
Since there have not been any detected frequency-dependent
effects in the position of the jet components (Britzen et al. in prep.),
we used 8 and 15 GHz data together for the derivation of
the trajectories of the jet components C0 and C1.

Around 1993, 9 years
after the ejection of the previous plasma component, a new
plasma component arises and follows the magnetic
tube in the beam, producing a new VLBI component
corresponding to C0 again (see Figure \ref{identification}).
In fact, the \textit{quasi-stationary component} C0 from Britzen
et al. (in prep.) corresponds to the quasi-stationary perturbation
of the magnetic tube in which the relativistic plasma propagates.
The magnetic tube deformation propagates with a speed $V_{a} \ll c$.

The existence of a quasi-stationary VLBI component,
which seems to be oscillating, can be explained by the fact that
firstly the magnetic perturbation of the beam
(magnetic loop) is mostly perpendicular to the ejection direction
(see Figure \ref{fig:Traj}) and secondly the observed time between
two ejections of the plasma component from the nucleus corresponds
to the time for a plasma component to cross the abovementioned
magnetic loop.

As the ejection direction is west, the magnetic
tube deformation propagating with the speed $V_{a} \ll c$ will
produce a slow motion in the west direction. Moreover, VLBI
observations of all the components show a slow bending to
the south after the C0 and C1 components. This slow bending
can be an effect of the slow motion of the BBH system around the
gravity center of the galaxy.

For the VLBI component Ca at $\sim 1.4$ mas, which is
situated just after C1 (see Figure
\ref{identification}) slow motion in the south-west direction has been
detected over the 20 years of observations. The fit of the model to
the trajectories of C0 and C1 predicts that this
component will be observed at the present position of Ca
and it will move slowly outwards (see Figure \ref{fig:Traj}). This
slow motion is due to both the slow motion of the magnetic
perturbation and of the BBH system around the gravity center of the
galaxy.


Since the information on the absolute position of the core is
lost in interferometric observations, we are not able to determine
whether the core is stationary or is itself moving. We will
investigate the stationarity of the core in phase reference
observations.

As indicated in the introduction, the VLBI core
will move due to:
\begin{itemize}
    \item the motion of the black hole ejecting the VLBI
    component around    the gravity center of the BBH system and
    \item the motion of the BBH system around the gravity center
    of the galaxy.
\end{itemize}

\subsection{Determination of the minimum of $\chi^{2}(i_{o})$}
\label{sec:minimum_io}

We first assume $M_{1}=M_{2}$, i.e. that the 2 masses of the
BBH system are equal.

From the shape of the trajectory, we find that the precession
is defined by $-\omega_{p}(t-z/V_{a})$, thus the orbital rotation
has to be $\omega_{b}(t-z/V_{a})$.

As indicated previously, to find the minimum of $\chi^{2}(i_{o})$,
we minimize $\chi^{2}_{t}(i_{o})$ as the inclination angle varies
gradually between two values. At each step of $i_{o}$, we determine
each free parameter $\lambda$ such that $\chi^{2}_{t}(\lambda)$ is
minimal for $\lambda$.

To determine the trajectory, we integrate (\ref{eq:zc}) using a step
$\Delta t$. By calling $\sigma_{c}$ the size of the ejected component,
the parameter $n_{rad}$, describing the length of the VLBI component
along the beam, can be related to $\sigma_{c}$ as follows

\begin{eqnarray}
    \sigma_{c} = n_{rad} \: sin(i_{o}) \: c \:\Delta t/(D_{a}mas) \ .
    \label{eq:sigma_c}
\end{eqnarray}

As the inclination varies, we change the integration step in order
to keep $\sigma_{c} = Cst$ if $n_{rad} = Cst$.

As the 2 components C0 and C1 are within 1 $mas$ and the
trajectory is complicated, the direction of the ejection, i.e. the parameter
$\Delta\Xi$, and the damping time, $T_{d}$, of the beam perturbations
cannot be properly constrained by the fit of the C0 and C1 coordinates.
To find the solution
\begin{itemize}
    \item the parameter $\Delta\Xi$ is maintained constant, i.e.
    $\Delta\Xi = 271^{o}$,
    \item the parameter $T_{d}$ is kept as a free parameter but with
    a maximal value $T_{d} \leq T_{d,max} = 1000$ $yrs$
    (\footnote{The value $T_{d,max} = 1000$ $yrs$ has been chosen to
    limit the expansion of the perturbation after the third VLBI
    component and values $T_{d} \geq T_{d,max}$ do not significantly change
    the value of $\chi^{2}$}).
\end{itemize}

\begin{figure}[ht]
\centerline{
\includegraphics[scale=0.5, bb =-200 -20 700 350,clip=true]{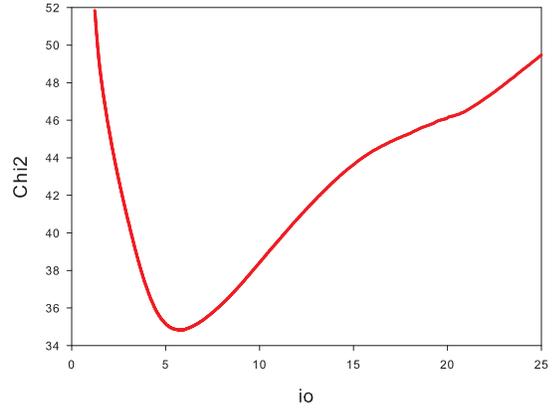}}
\caption{The variations of $\chi^{2}_{t}(i_{o})$. There is only one minimum in this
interval and it is located at $i_{o} \approx 5.8^{o}$. The fit has been
made using 66 points. As the model uses 12 free parameters the solution is
well constrained, as shown by the variations of the curve $\chi^{2}(i_{o})$}.
\label{fig:chi2_io}
\end{figure}

We calculated $\chi^{2}_{t}(i_{o})$ when the inclination angle
varies between $1^{o}$ and $25^{o}$. The result is shown in
Figure \ref{fig:chi2_io}

The inclination angle corresponding to the minimum is $i_{o} \approx 5.8^{o}$ and the
determination of the $1\sigma$ using (\ref{eq:1_sigma}) provides
$(\Delta i_{o})_{1\sigma-}\approx 1.8$ and $(\Delta i_{o})_{1\sigma+}\approx 1.7$.
Thus we have

\begin{eqnarray}
    i_{o} = 5.8 \; ^{+1.7}_{-1.8} \ .
\end{eqnarray}

As we calculated the variations of $i_{o}$ step by step minimizing
$\chi^{2}$ for each variable at each step, we can deduce the range
of values corresponding to $1\sigma$ for each parameter when the
inclination varies between $i_{o} = i_{o,min} - (\Delta i_{o})_{1\sigma-}$
and $i_{o} = i_{o,min} + (\Delta i_{o})_{1\sigma+}$.

The bulk Lorentz factor of the VLBI component producing C0 and C1 is

\begin{eqnarray}
    \gamma_{c} = 3.7 \; ^{+0.3}_{-0.2} \ .
\end{eqnarray}

\begin{figure}[ht]
\centerline{
\includegraphics[scale=0.5, bb =-200 -20 700 350,clip=true]{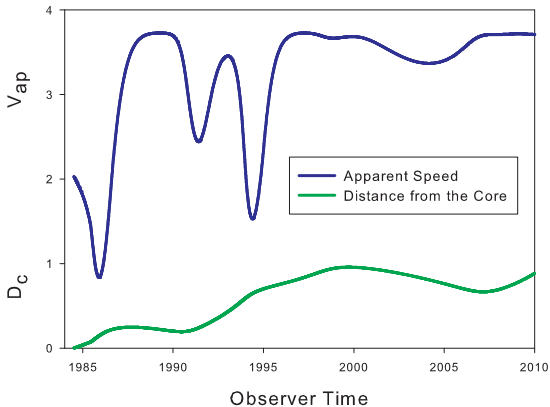}}
\caption{The variations with time of the apparent speed and
the distance from the core of the ejected component. We see that the
distance from the core can be mostly constant during certain periods
of time. During these periods, the apparent speed is high.}
\label{fig:V_ap-D_c}
\end{figure}

The variations of the apparent speed and the variations
of the distance from the core of the VLBI component are shown
in Figure \ref{fig:V_ap-D_c}. During the times when the motion of the component is mainly
perpendicular to the mean ejection direction, as the component moves
within a loop of the perturbated magnetic tube, its distance from the core
is mostly constant but its apparent speed is high, i.e. $v_{ap} \approx 3.7 \times c$.
The periods of time corresponding to these phases are between 1987 and 1990 for the
first one and around 2000 for the second one.

We will discuss the results of the BBH system parameters in the next section.
However, here we point out an important characteristic of the fit of 1807+784.
The better fits are obtained when the ratio $T_{p}/T_{b}$ has values close to
1, 2 or 3. The corresponding values of $\chi^{2}_{min}$ are given
in Table 1

\begin{center}
Table 1 : Values of $\chi^{2}_{min}$ for different ratios $T_{p}/T_{b}$ \medskip

\begin{tabular}
[c]{l||l}\hline
$T_{p}/T_{b}$           & $\chi^{2}_{min}$    \\\hline
$\approx 1$             & $47.1$              \\\hline
$\approx 2$             & $34.8$              \\\hline
$\approx 3$             & $37.6$              \\\hline
\end{tabular}
\end{center}

From Table 1, we see that the best fit is obtained when $T_{p}/T_{b} \approx 2.0$.
The results presented in this article correspond to this case.

In Table 2, we give the values of the geometric parameters characterizing the
trajectory of the VLBI component producing C0 and C1.

\begin{center}
Table 2 : Geometric parameters of the fit of C0 and C1 for $i_{o} \approx 5.8$\medskip%

\begin{tabular}
[c]{l||l|l|l}\hline
Parameter              & Value                        & Unit      & Remark        \\\hline
$\Delta\Xi$            & $271$                        & $^{o}$    & Fixed         \\\hline
$\Omega$               &$1.09\;^{+0.22}_{-0.24}$      & $^{o}$    &               \\\hline
$R_{o}$                &$581\;^{+0}_{-0}$             & $pc$      &               \\\hline
$T_{d}$                &$1000$                        & $yr$      &$T_{d,max}$    \\\hline
$t_{o}$                &$1984.51\;^{+0.32}_{-0.75}$   & $yr$      &               \\\hline
$n_{rad}$              &$165\;^{+24}_{-46}$           & $Steps$   &               \\\hline
$(\tau_{ejec})_{obs}$  &$0.96\;^{+0.33}_{-0.34}$      & $yr$      &Obs Frame      \\\hline
\end{tabular}
\end{center}

The parameter $R_{o}$ is found to be $R_{o} \gg 1$, which
indicates that the increase of the perturbation due to the
precession is in the linear regime, i.e. $R_{o} \propto z$
(see equation \ref{eq:Ro}). In that case, $\chi^{2}$ becomes
mostly independent of $R_{o}$.

The parameter $T_{d}$ reaches the maximum allowed value.

The ejection duration of the plasma responsible for the VLBI component
in the BBH system frame is
\begin{align}
    (\tau_{ejec})_{bbh} = n_{rad} \; \Delta t \approx 14.47\;yr\ ,
\end{align}
where $\Delta t$ is the integration step. Note that it
does not depend on the observing frequency.

We give in Table 2 the value of the duration of the ejection
of the VLBI component in the observer frame.
As the time compression factor due to the relativistic ejection
defined by $t_{obs}/t$ (see equation \ref{eq:tobs}) is $\approx 0.0668$,
the duration of the ejection of the VLBI component in the observer frame is
$(\tau_{ejec})_{obs} \approx 0.96$ $yr$.

The parameter $n_{rad}$ also characterizes the length of the VLBI
component along the beam. This length is $\sigma_{c} \approx 0.07$ $mas$.


The fit has been mqde using 66 points
(33 points for each coordinate). As the model uses 12 free
parameters (indeed $\Delta\Xi = Cst$ and $T_{d} = T_{d,max} = Cst$),
the solution is well constrained as shown by Figure \ref{fig:chi2_io}.


\subsection{Determination of the family of BBH systems}
\label{sec:family}

Starting from the solution found in section \ref{sec:minimum_io}, we fix
the parameters $i_{o} = i_{o,min}$, $t_{o} = 1984.51$, $n_{rad} = 165$ and
we study the variations of $\chi^{2}$ when the mass $M_{1}$ varies between
$10^{6}$ $M_{\odot}$ and $10^{11}$ $M_{\odot}$ (with the assumption $M_{1} = M_{2}$
still holding true).

This allows us to find whether, for a given mass, unique solutions,
or families of solutions exit. For the latter case, we find that
$\chi^{2}_{t} \approx Cst$ when the parameter $M_{1}$ varies, i.e. the parameter
$M_{1}$ shows a degeneration.

We find that there are 2 families corresponding to the ratio $T_{p}/T_{b} \approx 2.0$.
The results corresponding to these families are given in Table 3.

\begin{center}
Table 3 : Families of solutions with $T_{p}/T_{b} \approx 2.0$
for $i_{o} \approx 5.8$\medskip%

\begin{tabular}
[c]{l||l|l} \hline
    Family       &$\chi^{2}_{t}(M_{1})$ &$T_{p}/T_{b}$     \\\hline
$S1$             &$\approx 34.8$        &$1.967$            \\\hline
$S2$             &$\approx 35.8$        & $2.01$           \\\hline
\end{tabular}
\end{center}

The best fit corresponds to the family $S1$ for which we have

\begin{eqnarray}
    \frac{T_{p}}{T_{b}} \approx 1.967 \ .
\end{eqnarray}

To find the relations between the parameters of the family, we determined
the parameters for 4 different values of the mass $M_{1}$. The results are
given in Table 4 of Appendix II.

For all the \textit{members} of the family, the parameters of
Table 2 are the same and the parameters $\phi_{o}$ and $\psi_{o}$ are
also the same. Only the parameters $T_{p}$, $T_{b}$ and $V_{a}$
change when the mass changes. Moreover, they are correlated and
in Table 4 (Appendix II), the relations between them are given.

The precession period is proportional to the inverse of the square root
of the mass $M_{1}$, i.e.

\begin{eqnarray}
    T_{p} \propto M_{1}^{-1/2} \ ,
\end{eqnarray}
the ratio $T_{p}/T_{b}$ is the same for all the members of the family, so
\begin{eqnarray}
    T_{p} \approx 1.967 \; T_{b} \ ,
\end{eqnarray}
and
the Alfven speed is mostly proportional to the inverse of the precession
period, i.e.
\begin{eqnarray}
    V_{a} \propto 1/T_{p}^{0.954} \propto M_{1}^{0.477} \ .
\end{eqnarray}

Thus, if the parameters of a member of the family are known, we can
deduce the parameters for the rest of the members of the family, as $M_{1}$ varies.

From VLBI observations alone, we cannot determine the mass
of the BBH system in the nucleus of S5~1803+784. However from
the knowledge of the family of solutions, we can illustrate
the characteristics of the BBH system assuming a range of masses
corresponding, for instance, to the range between the solutions
$S1c$ and $S1d$, i.e. $7.14 \; 10^{8} \leq M_{1} \leq 5.92 \; 10^{9}$
$M_{\odot}$. We find that the precession period is
$2557.5 \geq T_{p} \geq 892.0$ $yr$ and the orbiting period is
$1299.7 \geq T_{b} \geq 453.2$ $yr$.

When $M_{1}=M_{2}$, the size of the BBH system is the same for all the
members of the family and is $0.100$ $mas$.

\subsection{The families of solutions with $M_{1} \neq M_{2}$}

In the previous sections we looked for solutions assuming $M_{1} = M_{2}$.

\begin{figure}[ht]
\centerline{
\includegraphics[scale=0.5, bb =-200 -20 700 350,clip=true]{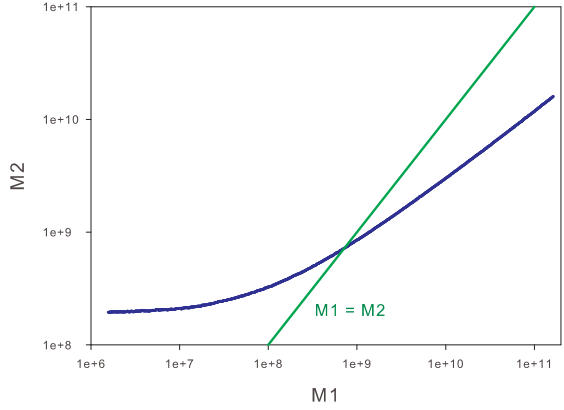}}
\caption{The family of possible BBH systems with $M_{1} \neq M_{2}$ corresponding
to the solution $S1c$ of Table 4 from Appendix~ ~II.}
\label{fig:M2_M1}
\end{figure}

However, it is possible to find solutions with $M_{1} \neq M_{2}$, starting
from the solutions found previously. To illustrate this possibility, we start from the
solution $S1c$ of the previous section (see Appendix II) and find the family of
solutions with $M_{1} \neq M_{2}$ which produces the same fit.

Assuming $i_{o} = 5.8^{o}$, $t_{o} = 1984.51$ and $n_{rad} = 165$, we gradually vary
the mass $M_{1}$ between $1.5 \; 10^{6}$ and $1.6 \; 10^{11}$ and we determine
$M_{2}$ minimizing $\chi^{2}$ for each free parameter at each step. The corresponding
range of $M_{2}$ is $1.9 \; 10^{8} \leq M_{2} \leq 1.6 \; 10^{10}$ and the corresponding
ratio $M_{1} / M_{2}$ is $0.01 \leq M_{1} / M_{2} \leq 10$.

The family has been plotted in Figure \ref{fig:M2_M1}.

\begin{figure}[ht]
\centerline{
\includegraphics[scale=0.5, bb =-200 -20 700 350,clip=true]{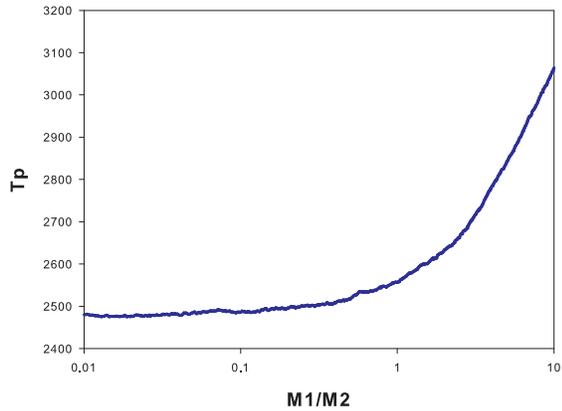}}
\caption{The precession period as a function of the ratio $M_{1} / M_{2}$.}
\label{fig:Tp_M1M2}
\end{figure}

The precession period corresponding to different ratios $M_{1} / M_{2}$ has been
plotted in Figure \ref{fig:Tp_M1M2}.

For all the \textit{members} of the family solution $S1c$ with $M_{1} \neq M_{2}$,
all the parameters of Table 2 are the same and the parameters $\phi_{o}$ and $\psi_{o}$
are also the same. Only the parameters $T_{p}$, $T_{b}$ and $V_{a}$
change when the ratio $M_{1} / M_{2}$ changes.

\begin{figure}[ht]
\centerline{
\includegraphics[scale=0.5, bb =-200 -20 700 350,clip=true]{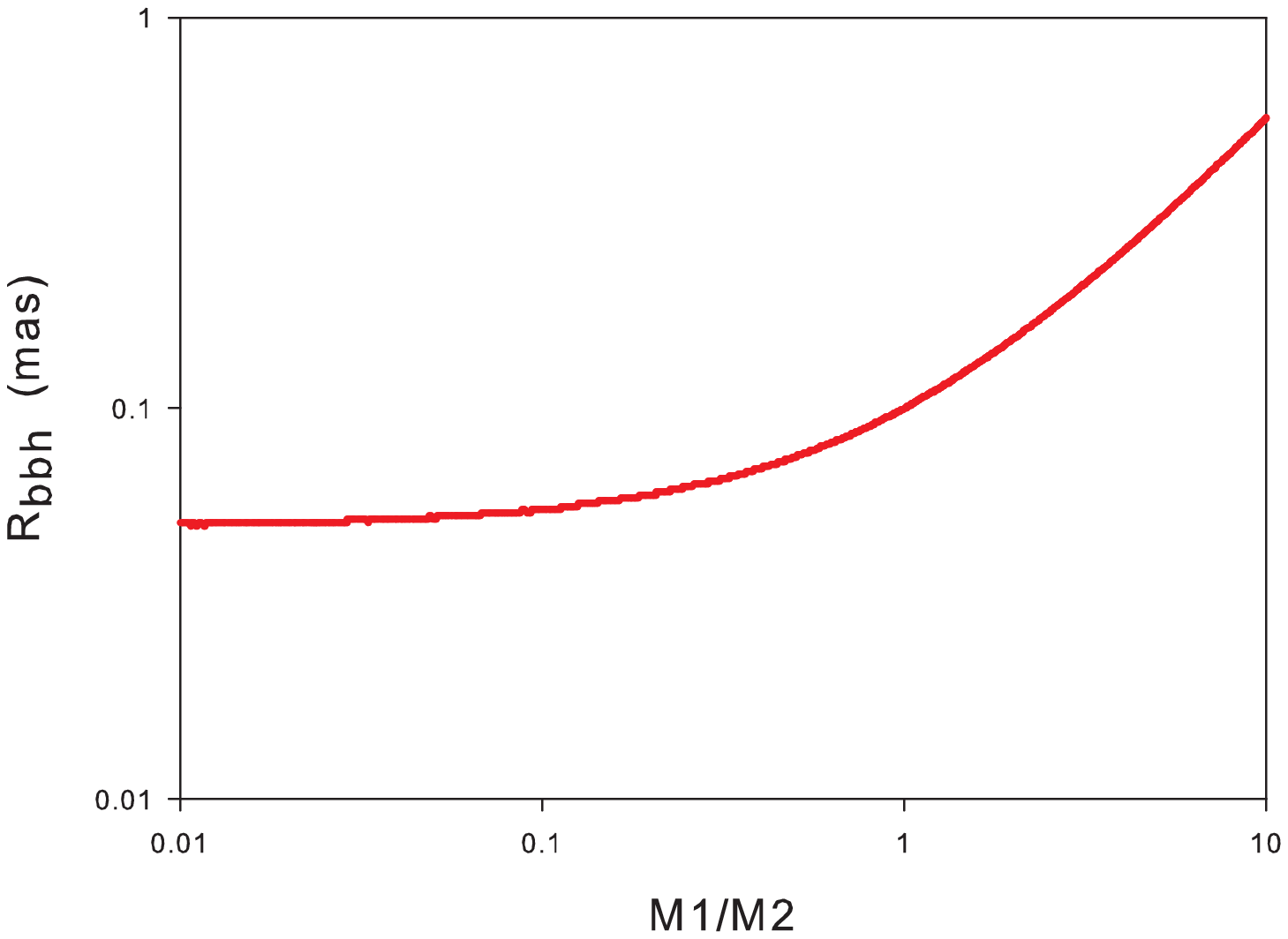}}
\caption{The size of the BBH system as a function of the ratio $M_{1} / M_{2}$.}
\label{fig:Rbbh_M1M2}
\end{figure}

The changes of the parameters $T_{p}$, $T_{b}$ and $V_{a}$ are characterized by
\begin{enumerate}
  \item when $M_{1} \rightarrow 0$, $M_{2} \rightarrow 1.98 \; 10^{8}$ $M_{\odot}$ ,
    \item when $M_{1} \ll M_{2}$, $T_{p}\approx Cst \approx 2480$ $yr$,
    \item when $M_{1} \gg M_{2}$, $T_{p}$ increases proportionaly with $M_{1} / M_{2}$,
    \item $T_{p} \approx 1.967 \; T_{b}$ ,
    \item the Alfven speed $V_{a}$ is mostly proportional to the inverse of $T_{p}$, i.e.
    \begin{eqnarray}
    V_{a} \propto 1/T_{p}^{0.958} \ ,
\end{eqnarray}
  \item the size of the BBH system varies from $R_{bbh} \approx 0.050$ $mas$
  when $M_{1} / M_{2} \rightarrow 0$ to $R_{bbh} \approx 0.558$ $mas$ when $M_{1} / M_{2} = 10$
  see Figure \ref{fig:Rbbh_M1M2}.
\end{enumerate}

Therefore, the main difference between the variations of the parameters $T_{p}$, $T_{b}$ and $V_{a}$
for families with $M_{1} = M_{2}$ and $M_{1} \neq M_{2}$ concerns the relation between
$T_{p}$ and $M_{1}$. When $M_{1} = M_{2}$, we have $T_{p} \propto M_{1}^{-1/2}$ and
when $M_{1} \neq M_{2}$ the variations between $T_{p}$ and $M_{1} / M_{2}$ are shown in
figure (\ref{fig:Tp_M1M2}).

\section{Discussion and Conclusion}

Assuming that the nucleus of a radio source contains a BBH system, we presented
a new method to fit the variations of both coordinates of a VLBI component as a function
of time. This method is self consistent and solves the problems of the previous
method, the two step method, used in Lobanov \& Roland (2005).
The presence of a BBH system produces two perturbations of the magnetic tube
in which the plasma, that is responsible for the radio emission of the VLBI component,
propagates. These two perturbations are due to the precession of the accretion disk
and the motion of the black holes around the gravity center of the BBH system.
The model is a geometric model and as we fit only the VLBI coodinates, the problem
reduces to an astrometric problem. As the knowledge of the variations of the
coordinates contains kinematical information, we are able to deduce
the inclination angle of the source and the bulk Lorentz factor of the ejected VLBI
component. The fit must be done following the trajectory
of one VLBI component and not using data of several components. Indeed, as a VLBI
component follows the perturbed magnetic tube, two VLBI components will not follow
exactly the same trajectories because the magnetic field perturbation propagates.
In addition to the two perturbations due to the BBH system, we can have a third one
due to the slow motion of the BBH system around the gravity center of the galaxy.
This third perturbation can be responsible for the slow bending of
the VLBI jet that is often observed in compact radio sources.
When it is observed, it is generally at a distance from the
core larger than a few $mas$. Then studying the motion of the VLBI component in
the first $mas$, we can expect that the influence of the slow motion of the BBH
system around the galactic gravity center is negligible compared to
the perturbations due to the precession of the accretion disk and the
motion of the black holes around the gravity center of the BBH
system. So to obtain the characteristics of the BBH system in the
nucleus of the radio galaxy, we have to fit the data concerning the
VLBI component within 1 or 2 $mas$ only.

To illustrate the method, we applied it to 1803+784 for which we have 20 years of
observations. This allows us to follow the VLBI component as it propagates
through the two first perturbations of the magnetic tube. These perturbations correspond to C0 and
C1, which have been called oscillatory VLBI components by observers.

Assuming, at first, that the two black holes have the same mass, we find
for 1803+784 that
\begin{enumerate}
    \item the inclination of the source is $i_{o} = 5.8 \; ^{+1.7}_{-1.8} \ ,$
    \item the bulk Lorentz factor of the ejected component is $ \gamma_{c} = 3.7 \; ^{+0.3}_{-0.2} \ ,$
    \item the angle between the accretion disk and the plane of the BBH system is
    $\Omega = 1.09\;^{+0.22}_{-0.24} \, $
    \item the size of the BBH system is $\approx 0.100$ $mas$ ,
    \item the precession period of the accretion disk ,$T_{p}$, and the orbiting period of the BBH
    system, $T_{b}$, are related by $T_{p} \approx 1.967 \; T_{b} \ ,$
    \item the origin of the component is $t_{o} = 1984.51\;^{+0.32}_{-0.75} \ ,$
    \item the duration of the ejection of the plasma responsible for the VLBI component is
    $(\tau_{ejec})_{obs} = 0.93\;^{+0.33}_{-0.34}$ $yr$ in the observer frame which corresponds
    to $(\tau_{ejec})_{bbh} \approx 13.9$ $yr$ in the BBH frame.
\end{enumerate}

We find that there is not a unique solution for the mass of the BBH system, but a
family of solutions that produces the same fit. For all the \textit{members}
of the family, all the parameters are the same, except the parameters
$T_{p}$, $T_{b}$ and $V_{a}$ which change when the mass changes. The variation laws
for these parameters are :
\begin{eqnarray}
    T_{p} \propto M_{1}^{-1/2} \ ,
\end{eqnarray}
\begin{eqnarray}
    V_{a} \propto 1/T_{p}^{0.954} \propto M_{1}^{0.477} \ ,
\end{eqnarray}
the ratio $T_{p}/T_{b}$ is the same for all the members of the family, so
\begin{eqnarray}
    T_{p} \approx 1.967 \; T_{b} \ ,
\end{eqnarray}
Thus, if the parameters of a member of the familly are calculated,
we can subsequently deduce the parameters of all the members of the
family when $M_{1}$ varies.

To illustrate the characteristics of the BBH system,
if we assume a range of masses corresponding to
$7.14 \; 10^{8}$ $M_{\odot}$ $\leq M_{1} \leq 5.92 \; 10^{9}$ $M_{\odot}$, we find that
the precession period is $2557.5$ $yr$ $\geq T_{p} \geq 892.0$ $yr$, the
orbit period is $1299.7$ $yr$ $\geq T_{b} \geq 453.2$ $yr$ .

Furthermore, from the knowledge of the solution with $M_{1} = M_{2}$,
we can find the families with $M_{1} \neq M_{2}$.

If VLBI observations allow us to find the possible
families of BBH system solutions and the main characteristics of these systems,
radio sources exist for which, in addition to VLBI data,
optical bursts associated with the birth of the VLBI component have been observed
(see for instance Britzen et al. 2001 in the case of 0420-014
and Lobanov \& Roland 2005 for 3C 345). The combination of VLBI and optical
observations can further constrain the families of solutions.

\begin{acknowledgements}
N. A. Kudryavtseva and M. Karouzos were supported for this research through a
stipend from the International Max Planck Research School (IMPRS) for
Radio and Infrared Astronomy at the Universities of Bonn and
Cologne. We thank the anonymous referee for carefully reading the
manuscript and many valuable comments.
\end{acknowledgements}

\newpage
\section{Appendix I : Coefficients A, B and C}
Let us call
\begin{equation}
\phi(t) = \omega_{p}t-k_{p}z(t)+\phi_{o}\ ,
\end{equation}
and
\begin{equation}
\psi(t) = \omega_{b}t-k_{b}z(t)+\psi_{o}\ .
\end{equation}

With
\begin{equation}
    x_1 = y_1 = -\frac{ M_{2}}{M_{1}+M_{2}}\left[\frac{T_{b}^{2}}{4\pi^{2}}G(M_{1}+M_{2})  \right]^{1/3},
\end{equation}
the coefficients $A$, $B$ and $C$ of equation (\ref{eq:dzt}) are given by

\begin{eqnarray}
    A & = & \exp(-2t/T_{d}) \left[ \frac{\omega_{b} \omega_{p} R(z)}{V_{a}^{2}} (y_{1} + x_{1}) \cos(\psi(t) - \phi(t))+ \right. \nonumber \\
    &&{}  \frac{\omega_{b}}{V_{a}} \frac{dR}{dz} (x_{1} + y_{1}) \sin(\psi(t) - \phi(t)) + \frac{\omega_{p}^{2} R(z)^{2}}{V_{a}^{2}} + \nonumber \\
    &&{}  \left. \left(\frac{dR}{dz}\right)^{2}  + \frac{\omega_{b}^{2}}{2 V_{a}^{2}} (x_{1}^{2} +   y_{1}^{2}) \right]  + 1 \ .
\end{eqnarray}

\begin{eqnarray}
    B & = & \exp(-2t/T_{d}) \left[ \frac{2 x_{1} \omega_{b} x_{1} \cos(\psi_{o})}{T_{d} V_{a}} \sin(\psi(t)) - \right. \nonumber \\
    &&{}  \frac{2 y_{1} \omega_{b} y_{1} \sin(\psi_{o})}{T_{d} V_{a}} \cos(\psi(t)) + \nonumber \\
    &&{}  2 \left(\frac{dR(z)}{dz} x_{1} \cos(\psi_{o}) - \frac{y_{1} \sin(\psi_{o}) R(z) \omega_{p}}{V_{a}}\right) \cos(\phi(t)) + \nonumber \\
    &&{}  2 \left(\frac{dR(z)}{dz} y_{1} \sin(\psi_{o}) + \frac{x_{1} \cos(\psi_{o}) R(z) \omega_{p}}{V_{a}}\right) \sin(\phi(t)) + \nonumber \\
    &&{}  \sin(\psi(t) - \phi(t)) \left\{ - \omega_{b} \frac{dR(z)}{dz}\left( x_{1} + y_{1} \right) + \right. \nonumber \\
    &&{}  \left. \frac{(\omega_{p} - \omega_{b})R(z)}{V_{a}}\left(  x_{1} + y_{1} \right)/T_{d} \right\} + \nonumber\\
    &&{}  \cos(\psi(t) - \phi(t)) \left\{- \frac{2 \omega_{p}\omega_{b}R(z)}{V_{a}}\left( x_{1}  + y_{1} \right) - \right. \nonumber \\
    &&{}  \left. \frac{dR(z)}{dz}\left( x_{1} + y_{1} \right)/T_{d} \right\} - \frac{\omega_{b}^{2}}{V_{a}}\left( x_{1}^{2} + y_{1}^{2}\right)\nonumber \\
    &&{}  \left.  - \frac{2 \omega_{p}^{2} R^{2}(z)}{V_{a}} -    \frac{2dR(z)}{dz}\frac{R(z)}{T_{d}}\right]\ .
\end{eqnarray}

\begin{eqnarray}
    C & = & \exp(-2t/T_{d})\left[ \frac{(x_{1} \cos(\psi_{o}))^{2} + (y_{1} \sin(\psi_{o}))^{2}}{T_{d}^{2}} + \right.\nonumber \\
    &&{}  2 \left(\frac{y_{1}\omega_{b}y_{1}\sin(\psi_{o})}{T_{d}} -  \frac{x_{1} x_{1}\cos(\psi_{o})}{T_{d}^{2}} \right) \cos(\psi(t)) - \nonumber \\
    &&{}  2 \left(\frac{x_{1}\omega_{b}x_{1}\cos(\psi_{o})}{T_{d}} -  \frac{y_{1} y_{1}\sin(\psi_{o})}{T_{d}^{2}} \right) \sin(\psi(t)) + \nonumber \\
    &&{}  2 R(z)\left(\frac{y_{1} \sin(\psi_{o})\omega_{p}}{T_{d}} -  \frac{x_{1} \cos(\psi_{o})}{T_{d}^{2}}  \right) \cos(\phi(t)) - \nonumber \\
    &&{}  2 R(z)\left(\frac{x_{1} \cos(\psi_{o})\omega_{p}}{T_{d}} +  \frac{y_{1} \sin(\psi_{o})}{T_{d}^{2}}  \right) \sin(\phi(t)) + \nonumber \\
    &&{}  \sin(\psi(t) - \phi(t)) \left\{ -  (\omega_{b} - \omega_{p}) R(z) \left( x_{1} + y_{1}\right)/T_{d} \right\} +  \nonumber \\
    &&{}  \cos(\psi(t) - \phi(t)) \left\{ \omega_{p} \omega_{b} R(z)\left( x_{1} + y_{1} \right) - \right. \nonumber \\
    &&{}  \left. R(z) \left( x_{1} + y_{1}\right)/T_{d}^{2} \right\} + \frac{\omega_{b}^{2}}{2}\left( x_{1}^{2} + y_{1}^{2} \right) +  R^{2}(z) \omega_{p}^{2} + \nonumber \\
    &&{}  \left. \left(R^{2}(z)+ \frac{x_{1}^{2}+y_{1}^{2}}{2} \right)/T_{d}^{2}\right] - v^{2}\ .
\end{eqnarray}

\newpage
\section{Appendix II : BBH parameters for 4 values of $M_{1}$ of the family S1}

The BBH parameters for 4 values of $M_{1}$ of the family S1 are given in Table 4.

\begin{center}
Table 4 : BBH parameters for 4 values of $M_{1}$ of the family S1\medskip%

\begin{tabular}
[c]{l||l|l|l|l}\hline
               & $S1a$           & $S1b$           & $S1c$           & $S1d$            \\\hline
$\phi_{o}$     & $78.9$          & $78.8$          & $79.0$          & $79.1$           \\\hline
$T_{p}$        & $16380.4$       & $6608.55$       & $2557.47$       & $891.95$         \\\hline
$M_{1}$        &$1.7556\;10^{7}$ &$1.0743\;10^{8}$ &$7.1437\;10^{8}$ &$5.9232\;10^{9}$  \\\hline
$\psi_{o}$     & $59.5$          & $59.5$          & $59.5$          & $59.8$           \\\hline
$T_{b}$        & $8326.9$        & $3360.29$       &$1299.68$        &$453.198$         \\\hline
$V_{a}$        & $0.0078855$     & $0.019298$      &$0.048358$       &$0.12665$         \\\hline
\end{tabular}
\end{center}

For all the \textit{members} of the family, all the parameters of Table 1 are the same
and the parameters $\phi_{o}$ and $\psi_{o}$ are also the same. Only the parameters
$T_{p}$, $T_{b}$ and $V_{a}$ change when the mass $M_{1}$ changes.

\end{document}